\documentclass[doublecol]{epl2} 
\usepackage{amsmath}

\title{Multifractal Height Cross-Correlation Analysis: A New Method for Analyzing Long-Range Cross-Correlations}
\shorttitle{Multifractal Height Cross-Correlation Analysis} 

\author{Ladislav Kristoufek\inst{1,2}}

\institute{                    
  \inst{1} Institute of Economic Studies, Charles University, Opletalova 26, Prague, CZ-110 00, Czech Republic\\
  \inst{2} Institute of Information Theory and Automation, Academy of Sciences of the Czech Republic, Pod Vodarenskou vezi 4, Prague, CZ-182 08, Czech Republic
}

\pacs{89.75.-k}{Complex systems}
\pacs{05.45.Tp}{Time series analysis in nonlinear dynamics}
\pacs{89.65.Gh}{Econophysics}

\abstract{
We introduce a new method for detection of long-range cross-correlations and multifractality -- multifractal height cross-correlation analysis (MF-HXA) -- based on scaling of $q$th order covariances. MF-HXA is a bivariate generalization of the height-height correlation analysis of Barabasi \& Vicsek \textit{[Barabasi, A.L., Vicsek, T.: Multifractality of self-affine fractals, Physical Review A 44(4), 1991]}. The method can be used to analyze long-range cross-correlations and multifractality between two simultaneously recorded series. We illustrate a power of the method on both simulated and real-world time series.}

\begin{document}

\maketitle

The research of long-range dependence and multifractality has been growing significantly in recent years with application to a wide range of disciplines \cite{Shiogai2010,Liao2010,Zuo2009,DiMatteo2005,DiMatteo2007,Chen2008,Rehman2009,Hayakawa2004,Barunik2010,Kristoufek2010}. Recently, the examination of long-range cross-correlations has become of interest as it provides additional information about the examined processes. Carbone \cite{Carbone2007} generalized the detrending moving average (DMA) method for higher dimensions. Podobnik \& Stanley \cite{Podobnik2008} adjusted the detrended fluctuation analysis for two time series and introduced the detrended cross-correlation analysis (DCCA). Zhou \cite{Zhou2008} further generalized the method and introduced the multifractal detrended cross-correlation analysis (MF-DXA). Jiang \& Zhou \cite{Jiang2011} then implemented moving average filtering to MF-DXA algorithm creating MF-X-DMA. In this paper, we introduce two new methods for an analysis of long-range cross-correlations -- the multifractal height cross-correlation analysis (MF-HXA) and its special case of the height cross-correlation analysis (HXA).

To analyze long-range cross-correlations, we generalize the $q$-th order height-height correlation function for two simultaneously recorded series. Let us consider two series $X_t$ and $Y_t$ with time resolution $\nu$ and $t=\nu,2\nu,...,\nu\lfloor\frac{T}{\nu}\rfloor$, where $\lfloor \rfloor$ is a lower integer sign. For better legibility, we denote $T^{\ast}=\nu\lfloor\frac{T}{\nu}\rfloor$, which varies with $\nu$, and we write the $\tau$-order difference as $\Delta_{\tau}X_t \equiv X_{t+\tau}-X_t$ and $\Delta_{\tau}X_tY_t \equiv \Delta_{\tau}X_t\Delta_{\tau}Y_t$. Height-height covariance function is then defined as

\begin{equation}
\label{eq1}
K_{xy,q}(\tau)=\frac{\nu}{T^{\ast}}\sum_{t=1}^{T^{\ast}/\nu}|\Delta_{\tau}X_tY_t|^{\frac{q}{2}} \equiv \langle|\Delta_{\tau}X_tY_t|^\frac{q}{2}\rangle
\end{equation}

where time interval $\tau$ generally ranges between $\nu=\tau_{min},\ldots,\tau_{max}$. Scaling relationship between $K_{xy,q}(\tau)$ and the generalized bivariate Hurst exponent $H_{xy}(q)$ is obtained as

\begin{equation}
\label{eq2}
K_{xy,q}(\tau) \propto \tau^{qH_{xy}(q)}.
\end{equation}

For $q=2$, the method can be used for the detection of long-range cross-correlations solely and we call it the height cross-correlation analysis (HXA). Obviously, MF-HXA reduces to the height-height correlation analysis of Barabasi \textit{et al.} \cite{Barabasi1991} for $X_t=Y_t$. Note that it makes sense to analyze the scaling according to Eq. \ref{eq2} only for detrended series $X_t$ and $Y_t$ and only for $q>0$ \cite{DiMatteo2007}. A type of detrending can generally take various forms --  polynomial, moving averages and other filtering methods -- and is applied for each time resolution $\nu$ separately. 

The bivariate Hurst exponent $0<H_{xy}(2)<1$ has similar properties and interpretation as a univariate Hurst exponent. For $H_{xy}(2)>0.5$, the series are cross-persistent so that a positive (a negative) value of $\Delta X_t \Delta Y_t$ is more statistically probable to be followed by another positive (negative) value of $\Delta X_{t+1} \Delta Y_{t+1}$. Conversely for $H_{xy}(2)<0.5$, the series are cross-antipersistent so that a positive (a negative) value of $\Delta X_t \Delta Y_t$ is more statistically probable to be followed by a negative (a positive) value of $\Delta X_{t+1} \Delta Y_{t+1}$. Note that even two pairwise uncorrelated processes can be cross-persistent\footnote{For example, let us have pairwise uncorrelated processes $X_t$ and $Y_t$ following fractional Gaussian noise with $H_{x}(2)=0.9$ and $H_y(2)=0.7$ and thus $H_{xy}(2)=0.8$. Even though the two processes are independently generated (and thus uncorrelated), they are cross-persistent. If $X_t>0$ and $Y_t>0$, then it is statistically more probable (based on persistence of the separate processes) that also $X_{t+1}>0$ and $Y_{t+1}>0$ than otherwise. Therefore, if the processes moved together in period $t$, it is statistically more likely that they will move together in period $t+1$ as well (and vice versa), i.e. the processes are cross-persistent.}.

The expected values of the bivariate Hurst exponents have been partly discussed in \cite{Podobnik2008, Zhou2008,Jiang2011}. It has been shown that 

\begin{equation}
H_{xy}(q)=\frac{H_x(q)+H_y(q)}{2}
\label{eq3}
\end{equation}

for all $q>0$ for pairwise uncorrelated and correlated processes. We present some new insights into this relation. To better understand the behavior of the bivariate Hurst exponent, we use a standard multifractal formalism \cite{Calvet2008}. Consider processes $X_t$ and $Y_t$ are multifractal with generalized Hurst exponents $H_x(q)$ and $H_y(q)$ so that

\begin{equation}
\langle|\Delta_{\tau}X_t|^q\rangle\propto \tau^{qH_{x}(q)}
\label{eq4}
\end{equation}
\begin{equation}
\langle|\Delta_{\tau}Y_t|^q\rangle\propto \tau^{qH_{y}(q)}.
\label{eq5}
\end{equation}

In the same way, we can write the joint scaling of two series (compare with Eq. \ref{eq1}) as 
\begin{equation}
\langle|\Delta_{\tau}X_tY_t|^\frac{q}{2}\rangle\propto \tau^{qH_{xy}(q)}.
\label{eq6}
\end{equation}

Using the definition of covariance, the left part of Eq. \ref{eq6} can be rewritten as
\begin{equation}
\footnotesize
\langle|\Delta_{\tau}X_tY_t|^\frac{q}{2}\rangle = \langle|\Delta_{\tau}X_t|^\frac{q}{2}\rangle\langle|\Delta_{\tau}X_t|^\frac{q}{2}\rangle+cov(|\Delta_{\tau}X_t|^\frac{q}{2},|\Delta_{\tau}Y_t|^\frac{q}{2}).
\label{eq7}
\end{equation}

From Eqs. \ref{eq4} and \ref{eq5}, the first part of the right-hand side of Eq. \ref{eq7} implies
\begin{equation}
\langle|\Delta_{\tau}X_t|^\frac{q}{2}\rangle\langle|\Delta_{\tau}Y_t|^\frac{q}{2}\rangle \propto \tau^{q\frac{H_{x}(q)+H_{y}(q)}{2}}
\label{eq8}
\end{equation}
which corresponds to Eq. \ref{eq3}. Therefore, the crucial part of long-range cross-correlations and multifractality is the scaling of covariances between $|\Delta_{\tau}X_t|^\frac{q}{2}$ and $|\Delta_{\tau}Y_t|^\frac{q}{2}$ with varying $\tau$. Consider now a scaling exponent $\alpha(q)$ and a scaling relationship
\begin{equation}
cov(|\Delta_{\tau}X_t|^\frac{q}{2},|\Delta_{\tau}Y_t|^\frac{q}{2}) \propto \tau^{q\alpha(q)}.
\label{eq9}
\end{equation}

This leads us to three simple implications. If covariances do not scale with $\tau$, then Eq. \ref{eq3} holds. If the covariances scale with $\tau$, the other two are as follows:
\begin{equation*}
\alpha(q) \ne \frac{H_{x}(q)+H_{y}(q)}{2} \Rightarrow H_{xy}(q) \ne \frac{H_{x}(q)+H_{y}(q)}{2}
\end{equation*}
\begin{equation}
\alpha(q) = \frac{H_{x}(q)+H_{y}(q)}{2} \Rightarrow H_{xy}(q) = \frac{H_{x}(q)+H_{y}(q)}{2}
\label{eq10}
\end{equation}

We show that these relationships are indeed true for artificially generated processes later in the text. Therefore, we need to distinguish between two types of long-range cross-correlations: (i) long-range cross-correlations caused by long-range dependence of the separate series, and (ii) long-range cross-correlations caused by scaling of covariances between $|\Delta_{\tau}X_t|^\frac{q}{2}$ and $|\Delta_{\tau}Y_t|^\frac{q}{2}$.

In order to test validity of the method, we present results for several artificial series. In the analysis, we apply MF-HXA  with changing $\tau_{max}=5,\ldots,100$ and fixed $\tau_{min}=1$. In turn, we obtain the 99\% jackknife confidence intervals under an assumption of a normally distributed Hurst exponent with an unknown variance. The estimated Hurst exponent is then taken as a mean of the exponents based on the various $\tau_{max}$. This way, we can comment on the results with statistical power \cite{DiMatteo2007}. In the procedure, we apply filtering of a constant trend. We now turn to the artificial processes.

\begin{figure}[htbp]
\center
\begin{tabular}{llll}
(a)&(b)&(c)&(d)\\
\includegraphics[width=1.6in]{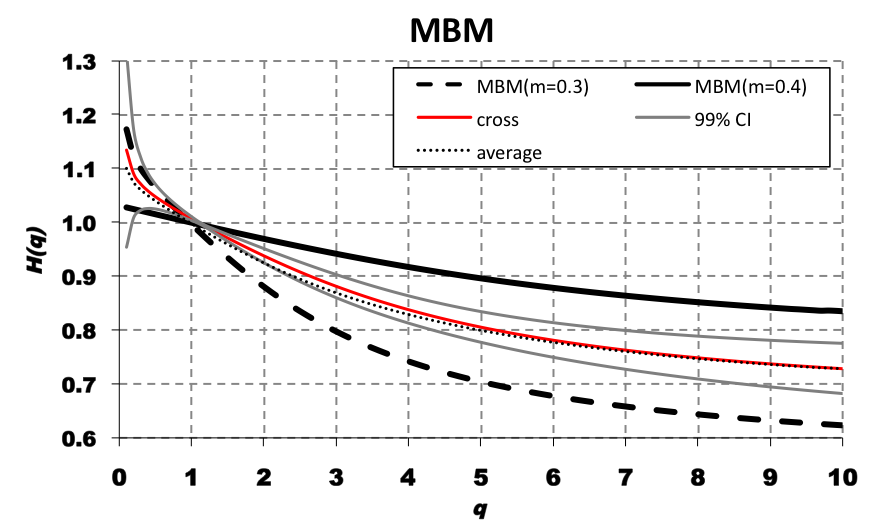}&\includegraphics[width=1.6in]{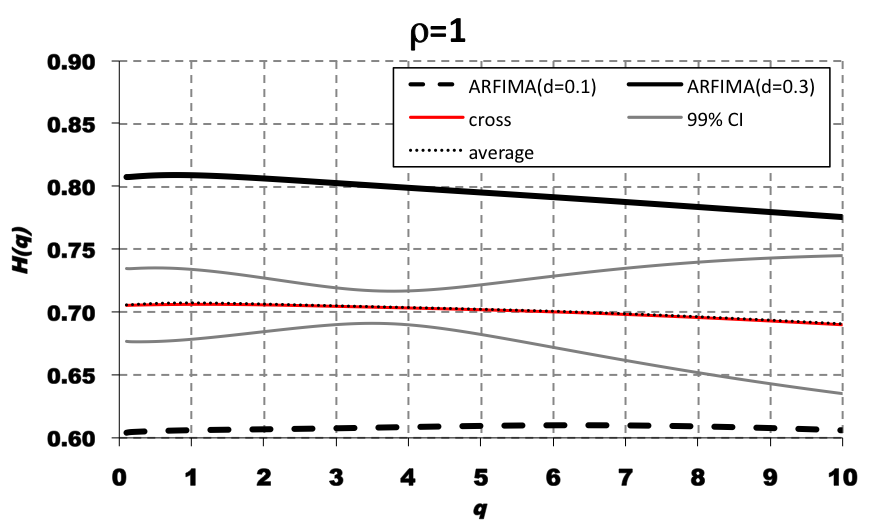}&\includegraphics[width=1.6in]{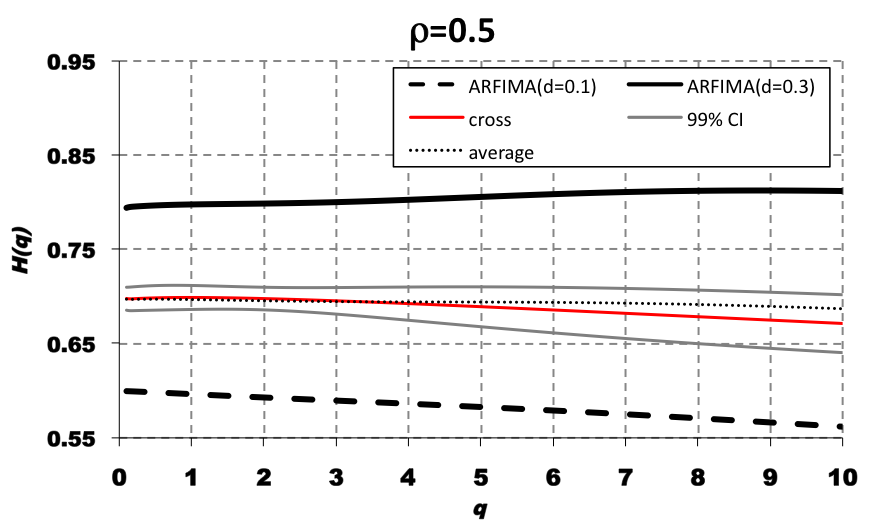}&\includegraphics[width=1.6in]{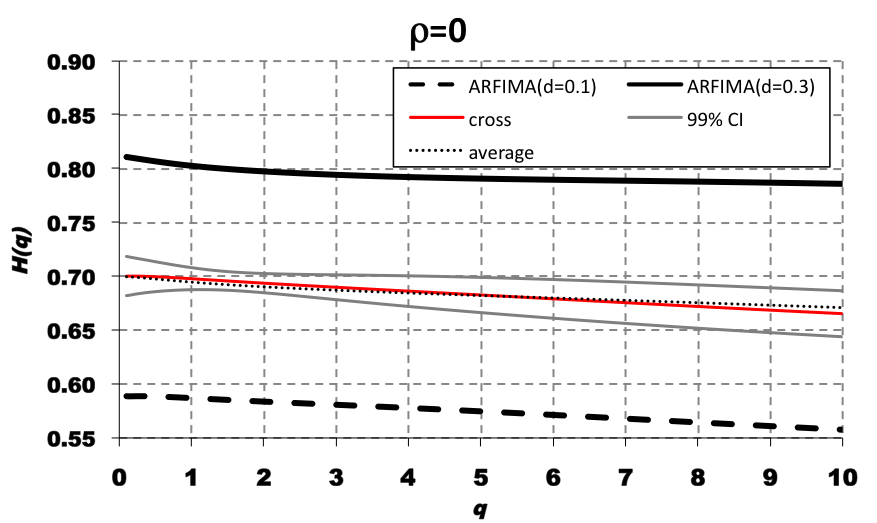}\\
(e)&(f)&(g)&(h)\\
\includegraphics[width=1.6in]{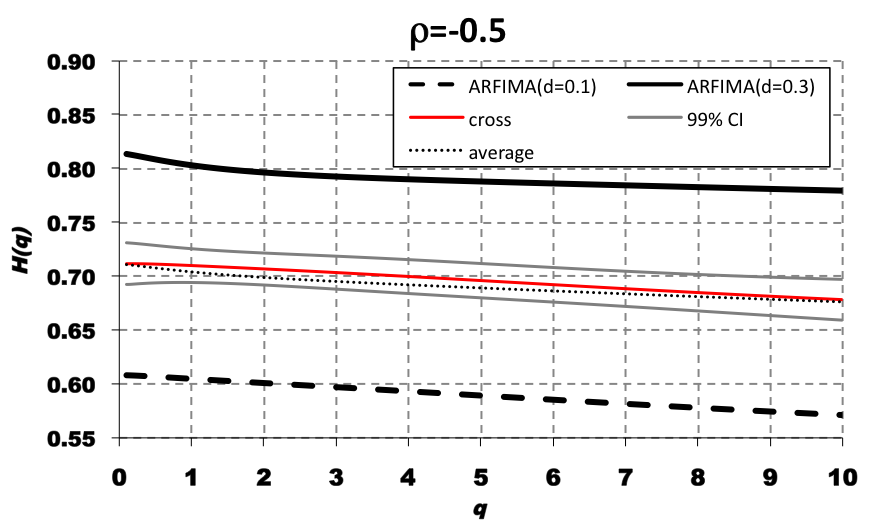}&\includegraphics[width=1.6in]{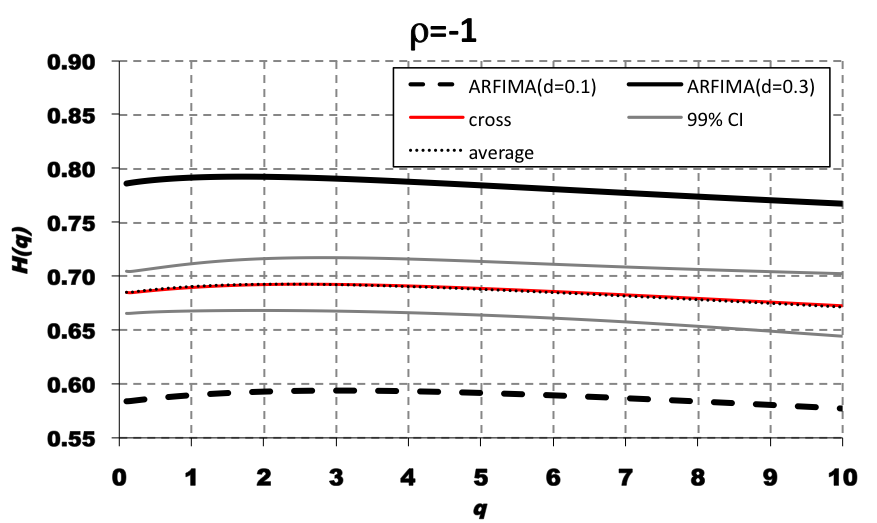}&\includegraphics[width=1.6in]{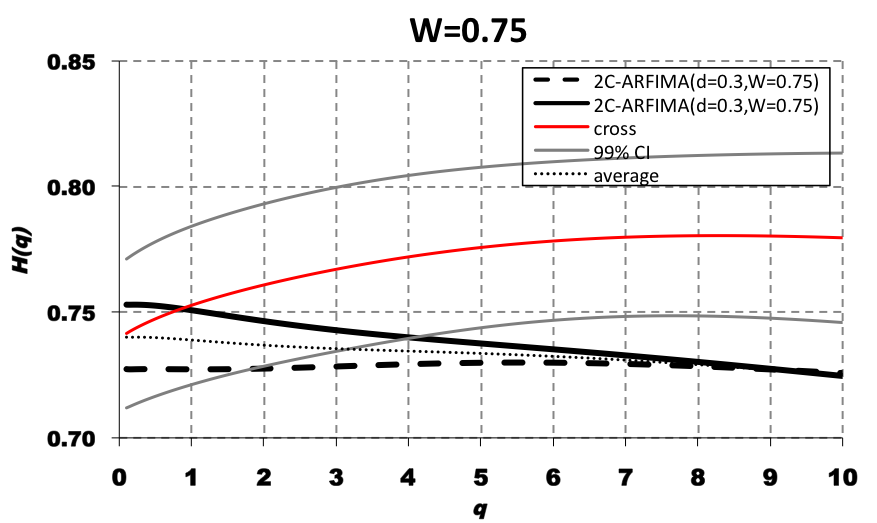}&\includegraphics[width=1.6in]{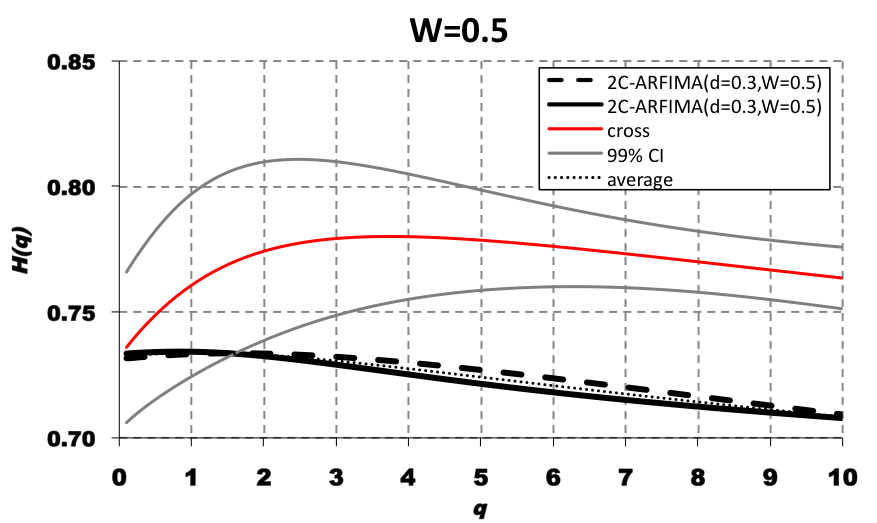}\\
\end{tabular}
\caption{\footnotesize\textit{(a) Binomial multifractal measures.} Generalized Hurst exponents (y-axis) dependent on moments $q\in\left[0.1,10\right]$ (x-axis) with step of 0.1. $H_x(q)$ for MBM with $m_0=0.3$ (bold black line) varies only weakly with $q$ compared to $H_y(q)$ for MBM with $m_0=0.4$ (bold dashed black line). $H_{xy}(q)$ (bold red line) is not significantly different from the average of $H_{x}$ and $H_{y}$ (dotted black line) for all $q$ (the 99\% jackknife confidence intervals around $H_{xy}(q)$ in gray). \textit{(b) - (f) ARFIMA(0,d,0) processes with correlated noise.} Generalized Hurst exponents (y-axis) dependent on moments $q\in\left[0.1,10\right]$ (x-axis) with step of 0.1. $H_x(q)$ for ARFIMA(0,d,0) with $H=0.8$ (bold black line) and $H_y(q)$ for ARFIMA($0,d,0$) with $H=0.6$ (bold dashed black line). The rest of the notation and parameters setting holds from \textit{(a)}. Figs. \textit{(b)} -- \textit{(f)} show ARFIMA($0,d,0$) processes with correlated noise with correlation coefficients $\rho=1$, $\rho=0.5$, $\rho=0$, $\rho=-0.5$ and $\rho=-1$, respectively. There is no significant deviation of $H_{xy}(q)$ from $\frac{H_x(q)+H_y(q)}{2}$ for all $q$ and for all examined correlations. \textit{(g) - (h) Two component ARFIMA($d_1,d_2$) processes.} Generalized Hurst exponents (y-axis) dependent on moments $q\in\left[0.1,10\right]$ (x-axis) with step of 0.1. Here, we use two component ARFIMA($d_1,d_2$) processes with $d_1=d_2=0.3$ and varying $W$. For $W=0.75$ \textit{(g)}, $H_{xy}(q)$ is significantly higher than $\frac{H_x(q)+H_y(q)}{2}$ for higher moments ($q>2.5$). For $W=0.5$ \textit{(h)}, the deviation of $H_{xy}(q)$ from $\frac{H_x(q)+H_y(q)}{2}$ is higher than for case (g) and the statistically significant deviation from the average starts at lower moments ($q > 1.3$). \label{fig1}}
\end{figure}

First, we start with the Mandelbrot's binomial multifractal (MBM) measures \cite{Meneveau1987,Mandelbrot1997}. Let $m_0>0$, $m_1>0$ and $m_0+m_1=1$ and let us work on interval [0,1]. In the first stage, the mass of 1 is divided into two subintervals [0,$\frac{1}{2}$] and [$\frac{1}{2}$,1], when there is the mass $m_0$ in the first subinterval and the mass $m_1$ in the second one. In the following stage, each subinterval is again halved and its mass is divided between the smaller subintervals in a ratio $m_0:m_1$. After $k$ stages, we obtain a series of $2^k$ values. Note that the values are deterministic as there is no noise added in the simplest version of the method. For an interval $[z,z+2^{-k}]$, the value $\mu$ has a value of $\mu[z,z+2^{-k}]=m_0^{k\varphi_0}m_1^{k\varphi_1}$, where $\varphi_0$ and $\varphi_1$ stand for the relative frequencies of numbers 0 and 1 in a binary development of $2^kz$, respectively. We construct two series with $m_0=0.3, 0.4$ and $k=16$. Results are presented in Fig. \ref{fig1}a, showing that the bivariate Hurst exponent $H_{xy}(q)$ does not deviate significantly from the average value of $H_x(q)$ and $H_y(q)$ even though the analyzed series are strongly correlated.

Second, we apply MF-HXA on ARFIMA processes with correlated noise terms. ARFIMA($0,d,0$) process is defined as $x_{t}=\sum_{i=1}^{\infty}{a_i(d)x_{t-i}+\varepsilon_t}$ where $0<d<0.5$ is a free parameter, related to Hurst exponent as $H=d+0.5$, and $a_i(d)=d\Gamma(i-d)/(\Gamma(1-d)\Gamma(1+i))$. We simulate long-range dependent series of length $10^4$. To describe influence of the correlations on MF-HXA estimates, we generate series with correlated noise $\varepsilon_t\sim N(0,1)$ and five cases are investigated -- correlation coefficients for the noise terms are set to 1, 0.5, 0, -0.5 and -1. The results are shown in Fig. \ref{fig1}b-f. The estimates of $H_{xy}(q)$ are not significantly different from $\frac{H_x(q)+H_y(q)}{2}$ for any $q$ or any correlation coefficient value. This result is in hand with the results shown in \cite{Jiang2011} -- pairwise correlations have no effect on the $H_{xy}$ estimation.

Third, we analyze the behavior of two-component ARFIMA processes \cite{Podobnik2008a}. For parameters $d_1$ and $d_2$, the two-component ARFIMA($d_1$,$d_2$) processes $X_t$ and $Y_t$ are described by the following set of equations:

\begin{center}$X_t=[Wx_t+(1-W)y_t]+\varepsilon_t$\\
$Y_t=[(1-W)x_t+Wy_t]+\nu_t$\\
$x_t=\sum_{i=1}^{\infty}{a_i(d_1)X_{t-i}}$\\
$y_t=\sum_{i=1}^{\infty}{a_i(d_2)Y_{t-i}}$\\
\end{center}
Here, $W$ is a free parameter ($0.5 \le W \le1$) controlling a strength of coupling between $X_t$ and $Y_t$, and $\varepsilon_t,\nu_t\sim N(0,1)$ are noise terms. Note that for $W=1$, we obtain two decoupled ARFIMA processes, whereas for $W<1$, the two processes have long memory of the process itself as well as of the other one. In our simulations, we consider $d_1=d_2=0.3$ with $W=0.5,0.75$ (practically, the case $W=1$ has been investigated in the previous paragraph). The results are shown in Fig. \ref{fig1}g,h. For both $W=0.75$ and $W=0.5$, we notice deviations of $H_{xy}(q)$ from $\frac{H_x(q)+H_y(q)}{2}$ starting already at $q=0.1$. The deviations are statistically insignificant for lower moments $q$ (due to rather short series, $T=10^4$), but become statistically significant for higher moments (for $q>1.3$ when $W=0.5$ and for $q>2.5$ when $W=0.75$). The effect gets stronger with lower $W$. Indeed, these are expected results as the construction of the two-component ARFIMA mixes the long memory of the separate processes together.

\begin{figure}[htbp]
\center
\begin{tabular}{llll}
(a)&(b)&(c)&(d)\\
\includegraphics[width=1.6in]{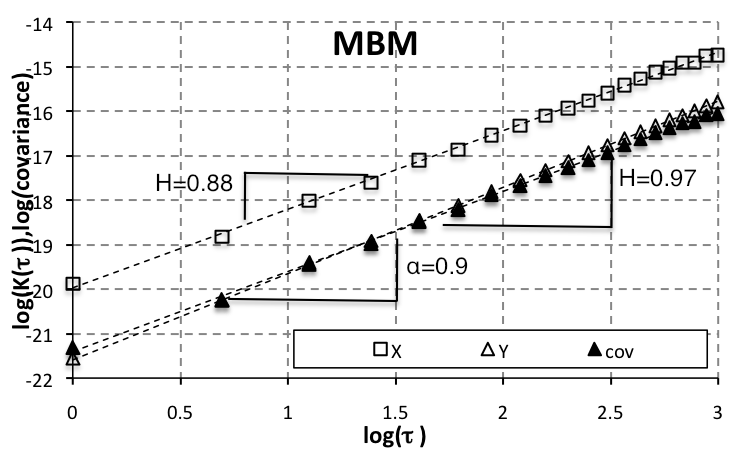}&\includegraphics[width=1.6in]{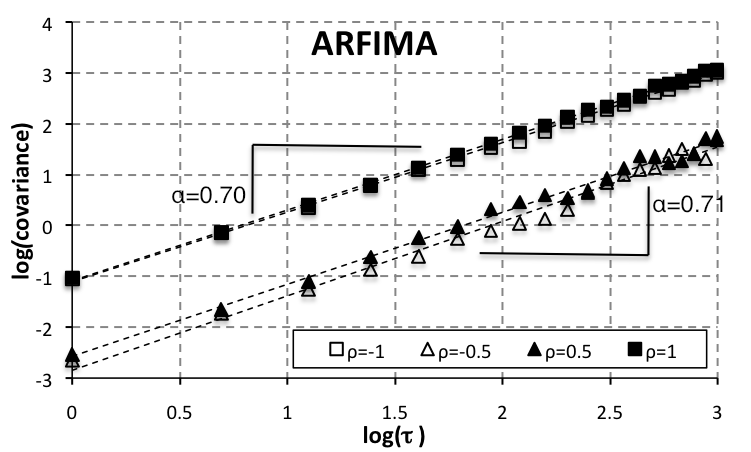}&\includegraphics[width=1.6in]{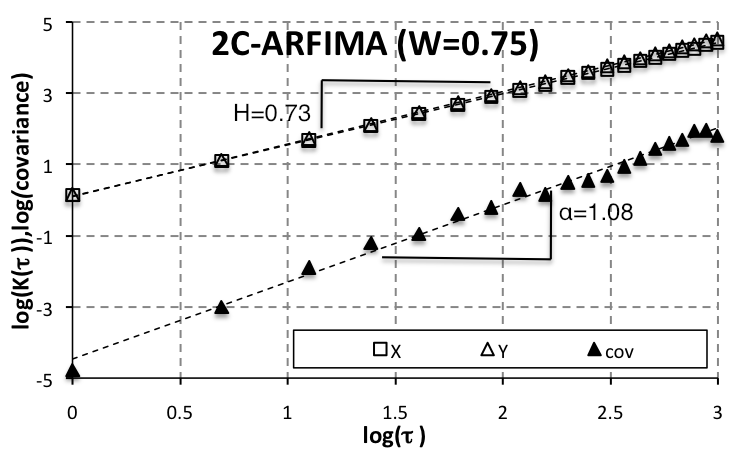}&\includegraphics[width=1.6in]{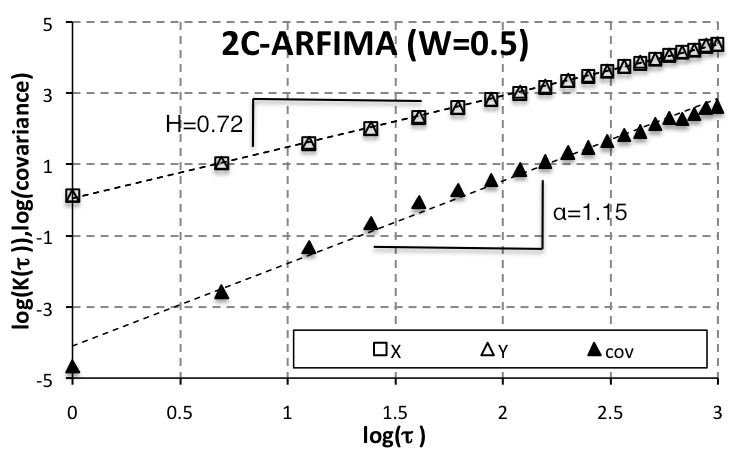}\\
\end{tabular}
\caption{\footnotesize\textit{Scaling of $K_{x,2}(\tau)$, $K_{y,2}(\tau)$ and covariances between $|\Delta_{\tau}X_t|$ and $|\Delta_{\tau}Y_t|$.} The scaling is shown for MBM \textit{(a)}, correlated ARFIMA processes \textit{(b)} and two-component ARFIMA processes \textit{(c,d)}. For MBM, we observe slight divergence of $\alpha(2)$ from the average of $H_x(2)$ and $H_y(2)$, which remains insignificant (see Fig. \ref{fig1}a). Correlated ARFIMA processes show practically perfect fit the expected $\alpha(2)$ of 0.7 (ARFIMA processes with $d=0.1$ and $d=0.3$). Two-component ARFIMA processes exhibit remarkable deviation of $\alpha$ from the average of Hurst exponents. Note that the fits (dashed black lines) and slopes are estimated on the whole sample from $\tau_{min}=1$ to $\tau_{max}=20$. The results are in hand with expectations based on Eqs. \ref{eq4} -- \ref{eq10} and in agreement with Fig. \ref{fig1}.}\label{fig2}
\end{figure}

In Fig. \ref{fig2}, we present the results based on separation in Eq. \ref{eq7}, i.e. scaling of separate processes and scaling of covariances of $|\Delta_{\tau}X_t|^\frac{q}{2}$ and $|\Delta_{\tau}Y_t|^\frac{q}{2}$. For illustrational purposes, we show only the case $q=2$. For MBM (Fig. \ref{fig2}a), the scaling of covariances is slightly lower than the average of Hurst exponents, yet remains well between them. This is reflected in the fact that the estimated $H_{xy}(2)$ is not equal to the average of estimated $H_x(2)$ and $H_y(2)$ but is rather close to the lower confidence interval (Fig. \ref{fig1}a), yet the deviation is still insignificant. In Fig. \ref{fig2}b, four cases of correlated ARFIMA processes are illustrated. All four processes show $\alpha(2) \approx 0.7$, which perfectly fits the expectations. We can see that the covariances are higher for highly correlated series than the less correlated series, but the scaling relation remains the same for all. The case of uncorrelated ARFIMA processes exhibits no scaling of covariances (as these vary around zero) and is thus not shown. In Figs. \ref{fig2}c,d, the two-component ARFIMA processes are illustrated. Here, the difference between scaling of covariances and the pair of $K_{x,2}(\tau)$ and $K_{y,2}(\tau)$ is remarkable for both $W=0.75$ and $W=0.5$. The scaling of covariances is expectedly stronger for $W=0.5$. These results perfectly support the calculations presented in Eqs. \ref{eq4} -- \ref{eq10}.

To show potential use of the method, we study different real-world financial series, which we consider the outputs of the complex systems -- daily volatility and volume series of NASDAQ and S\&P500 stock indices (finance.yahoo.com database), and daily returns and volatility of spot and futures prices of WTI Crude Oil (NYMEX Commodities database). Even though the real-world series are of the same length order as the simulated processes, which scale even up to $\tau=100$ and $q=10$, $K_{\tau}(q)$ usually does not scale for $\tau>20$ and $q>3$ for daily financial data \cite{DiMatteo2007}. Also, we apply linear filtering according to \cite{DiMatteo2005}. The generalized Hurst exponents are then estimated by varying $\tau_{max}$ between 5 and 20 for $0.1 \le q \le 3$. 

\begin{figure}[htbp]
\center
\begin{tabular}{lll}
(a)&(b)&(c)\\
\includegraphics[width=2.1in]{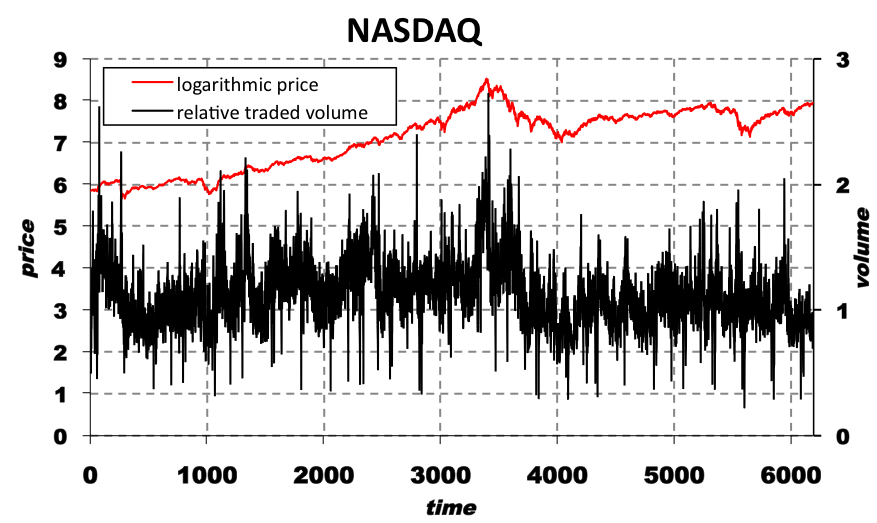}&\includegraphics[width=2.1in]{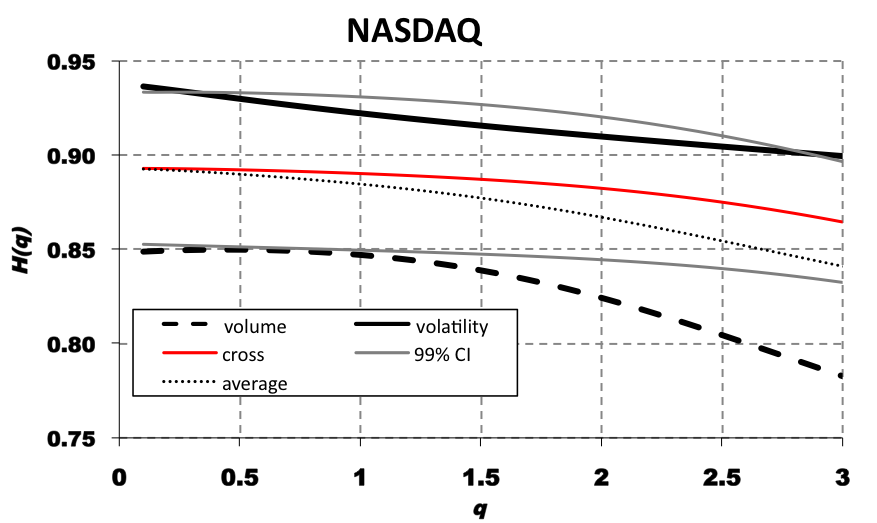}&\includegraphics[width=2.1in]{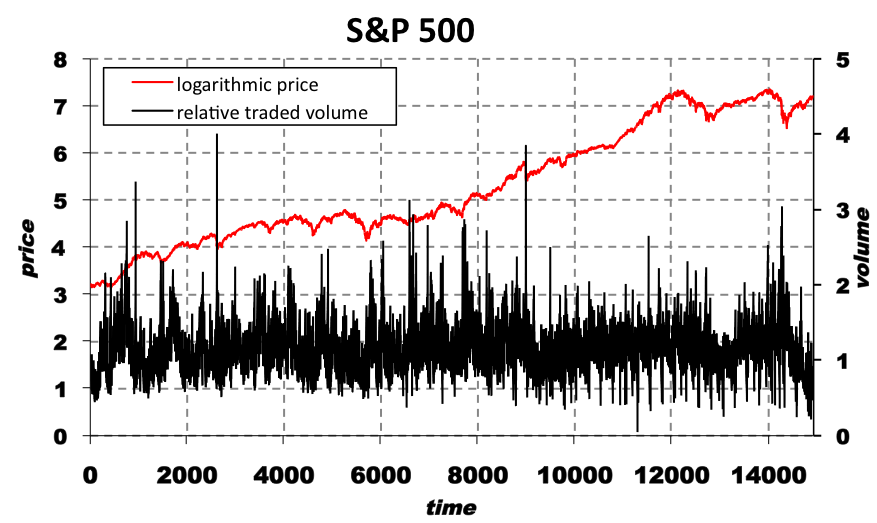}\\
(d)&(e)&(f)\\
\includegraphics[width=2.1in]{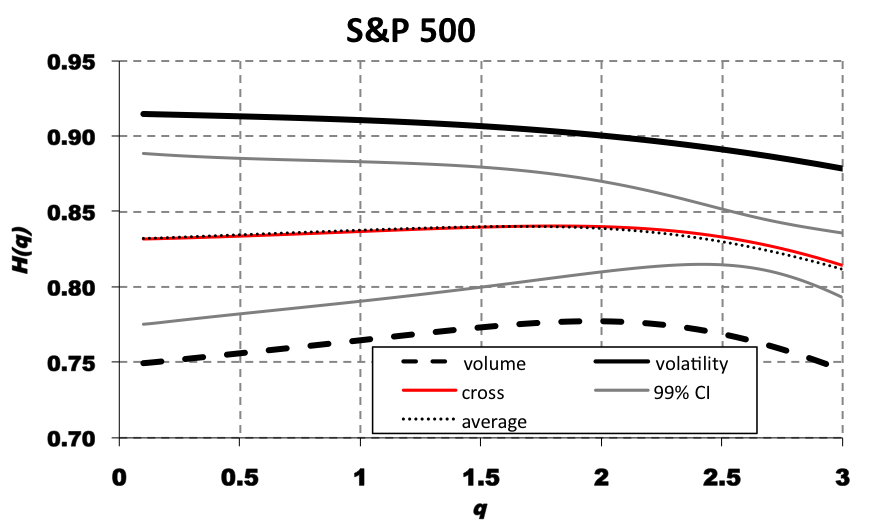}&\includegraphics[width=2.1in]{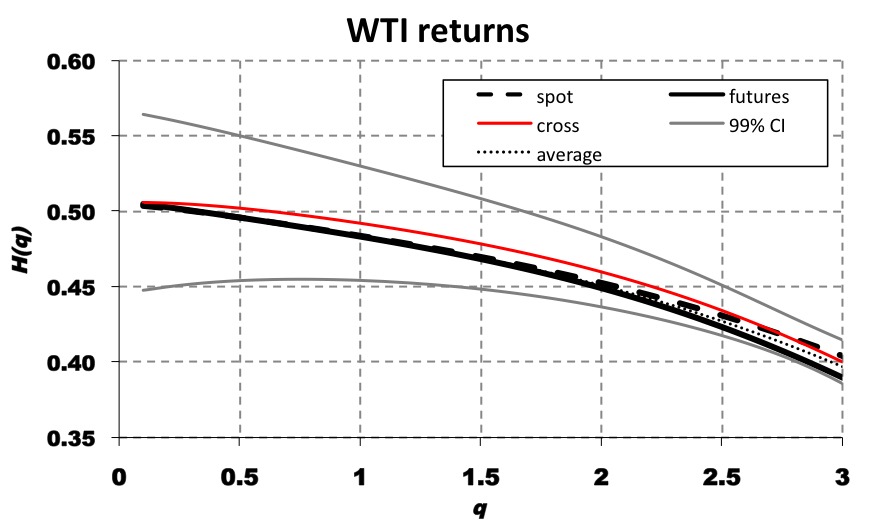}&\includegraphics[width=2.1in]{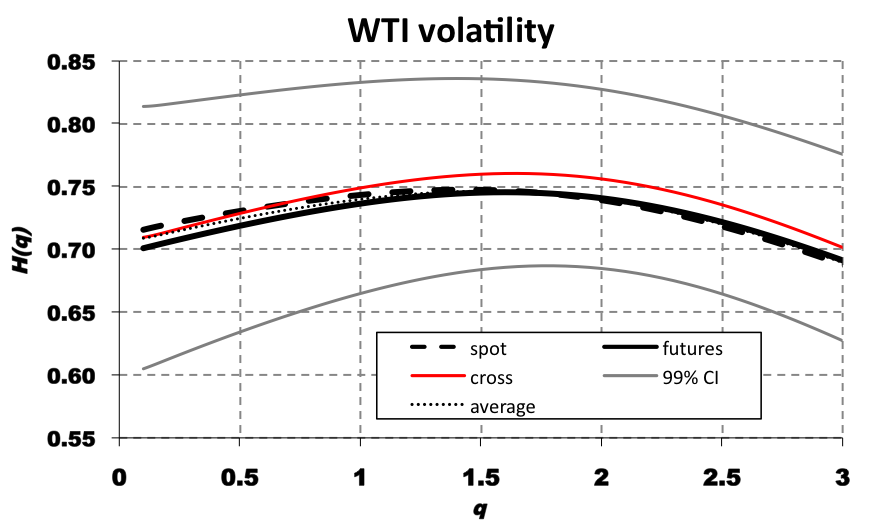}\\
\end{tabular}
\caption{\footnotesize\textit{(a), (b) NASDAQ.} (a) Evolution of logarithmic prices (red, left y-axis) and transformed traded volumes (black, right y-axis) in between 11.10.1984 and 26.4.2011. (b) Generalized Hurst exponents (y-axis) dependent on moments $q\in\left[0.1,3\right]$ (x-axis) with step of 0.1. $H_x(q)$ for NASDAQ volatility (bold black line) and $H_y(q)$ for NASDAQ traded volume (bold dashed black line) both vary with $q$ while stronger variation is present for volume. $H_{xy}(q)$ (bold red line) is not statistically different from the average (dotted line) of $H_x(q)$ and $H_y(q)$ for any $q$. \textit{(c), (d) S\&P500.} The time period covered ranges from 3.1.1950 to 26.4.2011. Same notation and estimation parameters setting hold here. $H_{xy}$ again does not differ from the average of the univariate Hurst exponents. \textit{(e) WTI crude oil spot and futures prices returns.} Same notation holds, $H_{x}$ represents the dynamics of spot returns and $H_{y}$ for futures returns. There is again no significant deviation of $H_{xy}(q)$ from $\frac{H_x(q)+H_y(q)}{2}$. \textit{(f) WTI crude oil spot and futures prices volatility.} The notation holds. Generalized Hurst exponents practically overlay for both series as well as for the joint dynamics.\label{fig3}}
\end{figure}

\begin{figure}[htbp]
\center
\begin{tabular}{llll}
(a)&(b)&(c)&(d)\\
\includegraphics[width=1.6in]{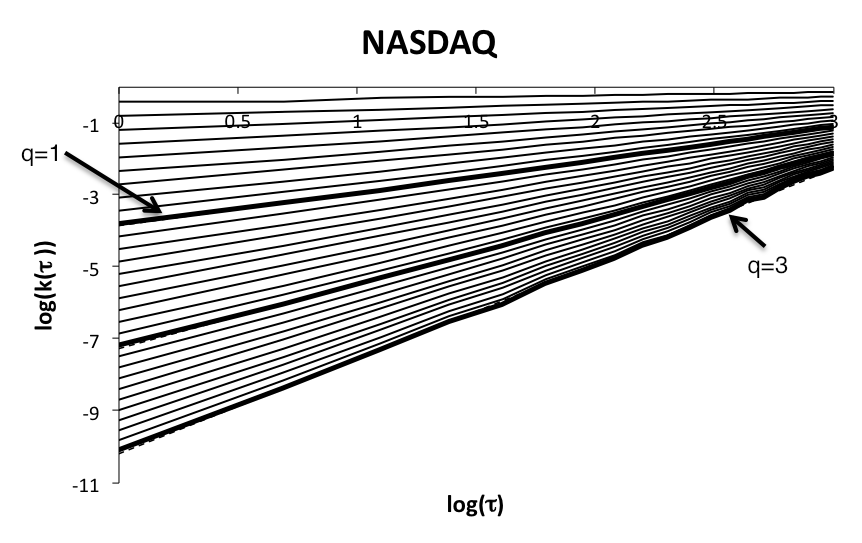}&\includegraphics[width=1.6in]{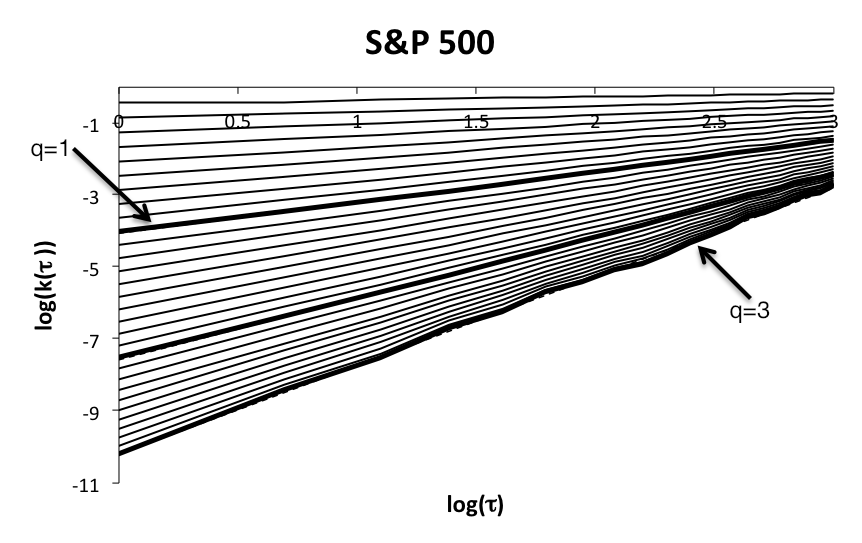}&\includegraphics[width=1.6in]{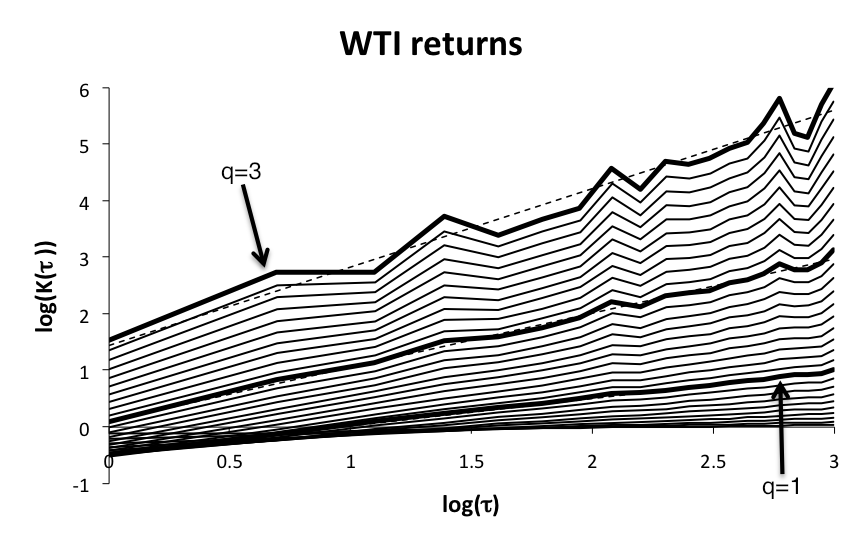}&\includegraphics[width=1.6in]{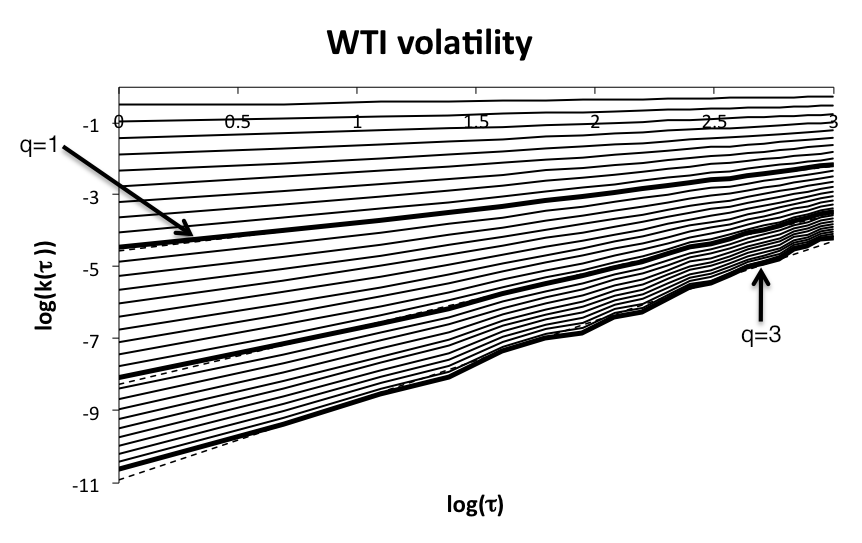}\\
(e)&(f)&(g)&(h)\\
\includegraphics[width=1.6in]{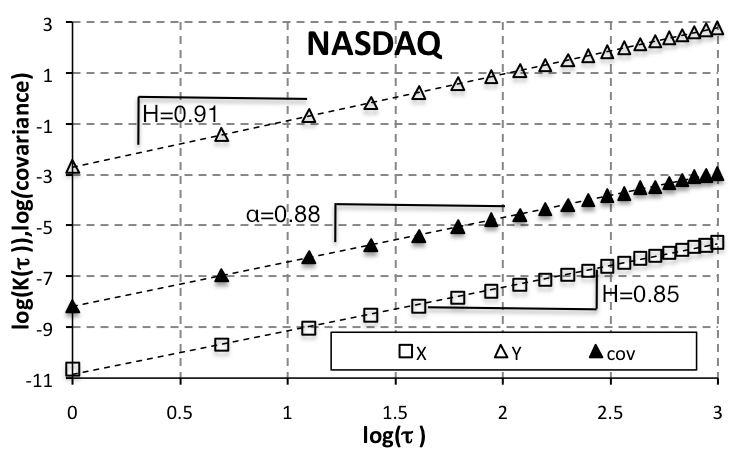}&\includegraphics[width=1.6in]{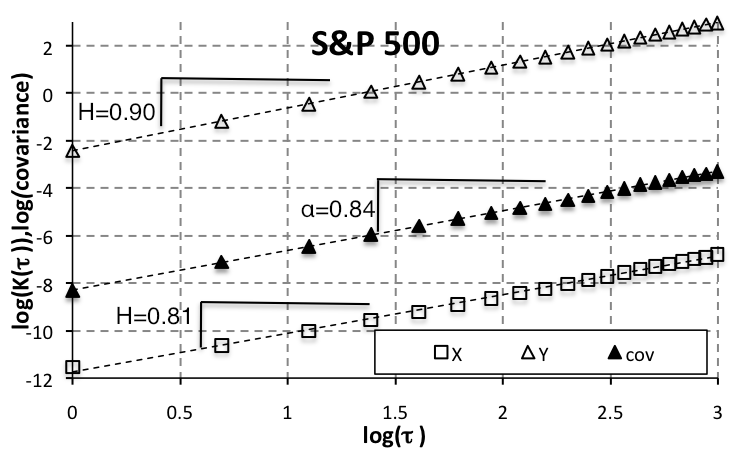}&\includegraphics[width=1.6in]{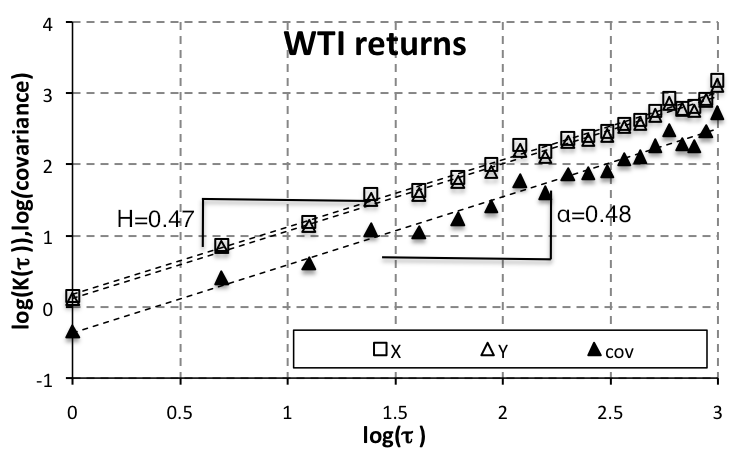}&\includegraphics[width=1.6in]{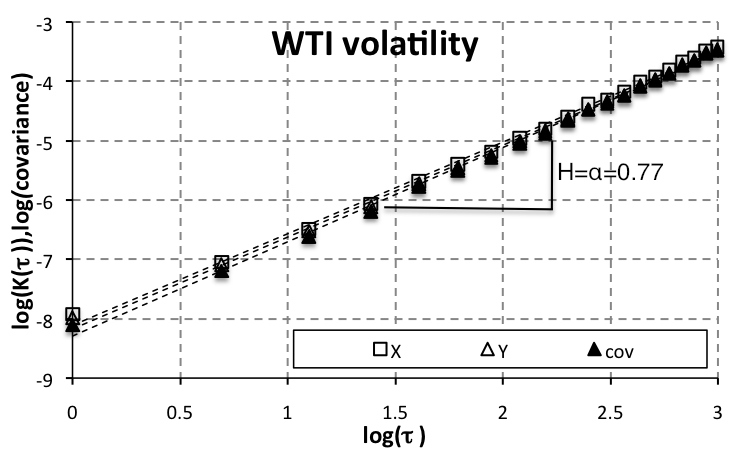}\\
\end{tabular}
\caption{\footnotesize\textit{(a) -- (d) Scaling of $K_{xy,q}$ for NASDAQ, SPX and WTI.} The scaling functions are presented by black lines, where the cases of $q=1,2,3$ are in bold. For these three cases, the best fits are illustrated (dashed lines). The scaling is very stable for NASDAQ, SPX and WTI volatility up to $\tau=20$ for all examined $q$ so that the fits are almost undistinguishable from the scaling functions. For WTI returns, the scaling is less stable with increasing $q$. For the analyzed series, it implies that scaling is better for higher values of Hurst exponents. \textit{(e) -- (h) Scaling of $K_{x,q}$, $K_{y,q}$ and covariances between $|\Delta_{\tau}X_t|$ and $|\Delta_{\tau}Y_t|$ for NASDAQ, S\&P500 and WTI.} Best linear fits are represented by dashed lines and estimated slopes are noted. For illustrational purposes, we show only the case $q=2$. The scaling exponents $\alpha(2)$ are approximately equal to the average of estimated Hurst exponents. This implies that eventual cross-persistence (for cases of NASDAQ,  S\&P500 and WTI volatility) is majorly caused by persistence of the separate processes and the fact that the processes are pairwise correlated. Note that the differences of estimates from Fig \ref{fig3} are caused by the fact that here, we estimate the exponents for $\tau$ between $\tau_{min}=1$ and $\tau_{max}=20$, while for Fig. \ref{fig3}, we use the jackknife estimates.\label{fig4}}
\end{figure}

For the stock indices, we analyze the series of volume and volatility for the longest available datasets -- from 11.10.1984 to 26.4.2011 for NASDAQ (6,693 observations) and from 3.1.1950 to 26.4.2011 for S\&P500 (15,428 observations). We take absolute returns, defined as $|\log P_t-\log P_{t-1}|$ where $P_t$ is a stock index closing price, as a measure of volatility. Volume series are transformed as a relative deviation from a moving average of traded volume in approximately past two trading years (500 observations) to control for an exponential increase of the traded volume in past decades (Fig. \ref{fig3}a,c). The estimated generalized Hurst exponents are shown in Fig. \ref{fig3}b,d. For both stock indices, the trading volume and volatility are strongly persistent as well as cross-persistent. Nevertheless, the bivariate Hurst exponent $H_{xy}(q)$ does not differ significantly from the average of $H_x(q)$ and $H_y(q)$, i.e. the cross-persistence of the series is mainly due to the persistence of the separate processes and the fact that the processes are correlated (Fig. \ref{fig4}e,f). The scaling of $K_{xy,q}(\tau)$ is very stable up to $\tau=20$ and for all examined $q$s (Fig. \ref{fig4}a,b). The results are in hand with \cite{Podobnik2008} who found weaker persistence of the process of traded volume. However, the definitions of traded volume differ from our study.

For the WTI spot and futures prices, we cover a period from 2.1.1986 to 26.4.2011 (6,348 observations) and analyze the logarithmic returns $r_t=\log(P_t)-\log(P_{t-1})$ and the volatility again in the form of absolute returns. The results are shown in Fig. \ref{fig3}e,f. For both returns and volatility, the estimates of the generalized Hurst exponents practically overlap for all $q$. On one hand, the returns show no signs of long-range correlations or cross-correlations. On the other hand, the volatility of separate processes show strong persistence as well as cross-persistence. Moreover, the generalized Hurst exponents vary only slightly with $q$ and are not even monotonically declining as expected for multifractal processes, suggesting that the processes of volatility are monofractal. Yet again, the cross-persistence of the series is mainly due to the persistence of the separate processes and high correlation between the processes (Fig. \ref{fig4}g,h) as $H_{xy}(q)$ does not significantly deviate from $\frac{H_x(q)+H_y(q)}{2}$. The scaling of $K_{xy,q}(\tau)$ shows different behavior for returns and volatilities. As for volatility, the scaling is very stable up to $\tau=20$ and $q=3$. On contrary, the scaling for returns becomes less stable with growing $q$ (Fig. \ref{fig3}c,d).

In conclusion, we introduce the new method for an analysis of long-range cross-correlations and multifractality -- the multifractal height cross-correlation analysis. The scaling of covariances of the absolute values of the series gives additional information about dynamics of two simultaneously recorded series and can cause divergence of the bivariate Hurst exponent from the average of the separate univariate Hurst exponents. A utility of the method has been shown on several artificial series as well as the real-world time series. We argue that even though majority of the analyzed series are cross-persistent, such cross-persistence is mainly caused by persistence of the separate processes and the fact that the series are correlated. The scaling of covariances of the absolute values of the examined processes is with good agreement with this result. A larger study comparing bias and efficiency of MF-HXA compared to the other methods analyzing long-range cross-correlations (MF-X-DFA and MF-X-DMA) shall follow.

\acknowledgments
The support from the Czech Science Foundation under Grants 402/09/H045 and 402/09/0965 and from the Grant Agency of the Charles University (GAUK) under project 118310 are gratefully acknowledged. The author would also like to thank J. Barunik and L. Vacha for helpful comments and discussions.

\bibliography{MFHXA}
\bibliographystyle{eplbib}

\onecolumn

\end{document}